\documentclass[aps,pra,twocolumn,groupedaddress,showpacs,floatfix]{revtex4}

\usepackage{color,soul}

\usepackage{epsfig}
\usepackage{amssymb}
\usepackage{amsmath}

\usepackage{rotfloat}
\usepackage{theorem}
\usepackage{epic}
\usepackage{graphics}
\usepackage{rotating}
\usepackage{placeins}
\usepackage{lscape}
\usepackage{longtable}
\usepackage{dcolumn}
\usepackage[usenames,dvipsnames]{pstricks}

\usepackage{epsfig}
\usepackage{hyperref}

\begin{document}

\newcolumntype{.}{D{.}{.}{1.5}}


\newcommand{\del}{\partial}
\newcommand{\beq}{\begin{equation*}}
\newcommand{\eeq}{\end{equation*}}
\newcommand{\be}{\begin{equation}}
\newcommand{\ee}{\end{equation}}
\newcommand{\beqa}{\begin{eqnarray}}
\newcommand{\eeqa}{\end{eqnarray}}
\newcommand{\bea}{\begin{eqnarray}}
\newcommand{\eea}{\end{eqnarray}}
\newcommand{\req}[1]{Eq.\,(\ref{#1})}

\newcommand{\bra}{\langle}
\newcommand{\ket}{\rangle}
\newcommand{\Tr}{{\rm Tr}\,}
\newcommand{\tr}{{\rm Tr}\,}
\newcommand{\s}{\sigma}
\newcommand{\w}{\omega}
\newcommand{\reci}[1]{\frac{1}{#1}}
\newcommand{\half}{\frac{1}{2}}
\newcommand{\emdash}{\hspace{1pt}---\hspace{1pt}}
\newcommand{\volint}[1]{\int \frac{d^4{#1}}{(2\pi)^4} \;}
\newcommand{\volthree}[1]{\int_0^{#1_F} \frac{d^3{#1}}{(2\pi)^3} \;}
\newcommand{\fixminus}{\raisebox{1.5pt}{\mathunderscore}\hspace{0.5pt}}
\newcommand{\etal}{{\it et al.}}

\newcommand{\putat}[3]{\begin{picture}(0,0)(0,0)\put(#1,#2){#3}\end{picture}}

\hypersetup{
    colorlinks=true,       
    linkcolor=red,          
    citecolor=green,        
    filecolor=magenta,      
    urlcolor=blue           
}

\urlstyle{same}


\title{Non-perturbative Analysis of the Influence of the Proton
  Magnetization and Charge Densities on the Hyperfine Splitting of
  Muonic Hydrogen}



\author{J.~D.~Carroll} 
\email{jcarroll@physics.adelaide.edu.au}
\author{A.~W.~Thomas}

\affiliation{Centre for the Subatomic Structure of Matter (CSSM),
School of Chemistry and Physics, University of Adelaide, SA 5005, Australia}
\homepage{http://www.physics.adelaide.edu.au/cssm}

\author{G.~A.~Miller}
\affiliation{University of Washington, Seattle, WA 98195-1560 USA}

\author{J.~Rafelski}
\affiliation{Departments of Physics, University of
  Arizona, Tucson, Arizona, 85721 USA}


\date{\today}

\begin{abstract}
We investigate the influence of the spatial extent of the proton
magnetization and charge densities on the 2{\it S} hyperfine splitting
in muonic hydrogen. The use of a non-perturbative relativistic Dirac
approach leads to corrections of 15\% to values obtained from the
perturbative treatment encapsulated by the Zemach radius, which
surpass the next-leading order contribution in the perturbation series
by an order of magnitude.
\end{abstract}

\maketitle

%
Precise values of the hyperfine splitting of the energy levels of
electronic and muonic hydrogen have long been known to be relevant to
precision tests of quantum
electrodynamics~\cite{RevModPhys.57.723,Eides:2000xc}. Interest in the
$2S$ state of muonic hydrogen has been strongly stimulated by the
recent precision measurement of the Lamb shift transition energy
between the $2P_{3/2}^{F=2}$ and $2S_{1/2}^{F=1}$ states by
Pohl~\etal~\cite{Pohl:2010zz}. This experiment led to the stimulating
conclusion that the proton root mean square radius differs from the
previously accepted value in the literature by 4.9
$\sigma$. Determining the proton radius depended on extracting the
Lamb shift of interest from the energy of the measured transition
between the $2S_{1/2}^{F=1}$ and $2P_{3/2}^{F=2}$ states. This
extraction relied on a variety of mainly perturbative calculations
performed by many authors, including those of
Martynenko~\cite{Martynenko:2004bt,Martynenko:2006gz},
Borie~\cite{Borie:2004fv,Borie:1982ax,Borie:1975xb},
Pachucki~\cite{Pachucki:1999zza,Pachucki:1996zza}, and many
others~\cite{Eides:2000xc,Rafelski:1977vq,Friar:1978wv,Fricke:1969fh}.
Our focus here is on the hyperfine splitting between the $2S$ states
with total spin $F=1$ and $F=0$. A computed value (22.8148(78)
meV~\cite{Martynenko:2004bt}) was used in ~\cite{Pohl:2010zz}. As a
guide to the importance of precision, we note that a change in this
value of 1\% would correspond to a shift by 1$\sigma$ in the proton
radius. Furthermore, this transition is under current experimental
investigation, with an announcement of a result expected soon.
 
%
%
The $F=0$ and $F=1$ levels of atomic $S$ states are split by the
interaction between the magnetic dipole moments of the lepton and
proton. Fermi computed this splitting treating the lepton and proton
as point-like particles having magnetic moments but no spatial extent,
and using the Coulomb potential to compute the non-relativistic
hydrogen wave function. This first-order treatment gives the Fermi
energy of the $nS$ states:
\be
E_F^{(nS)} = \frac{8}{3n^3}\alpha^4\frac{\mu_p m_\mu^2 M_p^2}{(m_\mu+M_p)^3},
\ee
where $\mu_p = 2.792847351$ is the proton magnetic moment, $\alpha$ is
the fine structure constant, and $M_p$, $m_\mu$ denote values of the
proton and muon masses. Zemach~\cite{Zemach:1956zz} included the
influence of the spatial extent of the proton charge and magnetization
distributions using first-order perturbation theory. The relevant
spin-spin interaction is given by
 \be \label{eq:Vzemach}
V_{\rm Zemach} =
\frac{8\pi\alpha \mu_p \vec{\s}_1\cdot\vec{\s}_2\rho_M(r)}
{12m_\mu M_p},
\ee
where the scalar product of the lepton and proton Pauli matrices,
$\vec{\s}_1\cdot\vec{\s}_2$ is +1 in the $F=1$ state and -3 in the
$F=0$ state, and $\rho_M$ is the magnetization density normalized to
unity. A contribution of the muon anomalous magnetic moment is
treated as a further (0.1$\%$) correction as per
Ref.~\cite{Martynenko:2004bt} to which we compare. Zemach treated the
difference between using $V_{\rm Zemach}$ and the corresponding
interaction obtained by taking the magnetization density to be a delta
function at the origin. His result can be expressed as
\bea
\nonumber
\Delta E_Z &=&
E_F\frac{2\mu\alpha}{\pi^2}\int\frac{d\vec{p}}{\vec{p\, }^4}
\left[\frac{G_E(-\vec{p\, }^2)G_M(-\vec{p\, }^2)}{\mu_p}-1\right] \\
&=& E_F(-2\mu\alpha)R_p\, ,
\label{eq:EZzemach}
\eea
where $R_p$ denotes the appropriately-named Zemach radius of
approximate value $R_p = 1.040(16)$~fm~\cite{Martynenko:2004bt} and
$\mu$ denotes the reduced muon mass. The appearance of $G_E$ results
from its influence on the bound-state wave function at the origin and
$G_M$ enters directly in the interaction. Effects of the spatial
extent of the magnetization density and higher order effects of
$G_M\ne1$ are not included.

Our aim here is to treat the effects of $V_{\rm Zemach}$ exactly
within the framework of very precise numerical solutions of the two
separate $F=0$ and $F=1$ Dirac equations. The dominant binding
interaction is taken to be the sum of the Coulomb potential and its
lowest-order correction arising from vacuum polarization, each being
modified by the non-zero spatial extent of the
proton~\cite{Carroll:2011rv}. Since the size of the hyperfine
splitting is much less than the size of the Lamb shift, a high degree
of numerical accuracy is required. The method we use to do this is
detailed in a previous publication~\cite{Carroll:2011rv}. These
high-precision calculations have been shown~\cite{Carroll:2011rv} to
reproduce exact, analytic eigenvalues for a point-Coulomb potential
and to satisfy a virial theorem test to within $0.5~\mu$eV which we
set as an upper bound on our numerical errors here.

To proceed we need to specify the magnetization density $\rho_M$ and
corresponding magnetic radius $(r_p^M)^2 \equiv \int
d^3r\;r^2\rho_M(r)$ (while including the effects of the charge radius
$(r_p^C)^2 \equiv \bra r_p^2\ket_C$ in the dominant binding
potentials). The magnetization density is taken to be a normalized
exponential
\be \label{eq:CD}
\rho_M(r) = \frac{\eta}{8\pi} e^{-\eta r}; \quad \eta =
\sqrt{12/\bra r_p^2\ket_M}\, .
\ee

The astute reader will question whether this exponential form is a
suitable choice for the magnetization density. While the dependence of
the Lamb shift energy on the shape of the charge distribution has been
fully investigated in Ref.~\cite{Carroll:2011de}\emdash and
discounted\emdash this conclusion does not automatically extend to the
magnetization density, which appears un-integrated in
Eq.~(\ref{eq:Vzemach}), as opposed to the integrated appearance in the
finite-Coulomb and finite vacuum polarization
potentials~\cite{Carroll:2011rv}. We shall return to this issue, but
for now an exponential form shall be assumed.

We calculate the converged eigenvalues for the $2S$ hyperfine
splitting (being the difference between the eigenvalues for the $F=0$
and $F=1$ hyperfine states) by numerically integrating the Dirac
equation 
\be
\label{eq:dirac}
\displaystyle{\frac{d}{dr}}\begin{pmatrix}G_{2S}(r)\\[2mm] F_{2S}(r)\end{pmatrix} 
= \left(\begin{array}{cc}
{\displaystyle -\frac{\kappa_{2S}}{r}} & \lambda_{2S} + 2\mu - V \\[2mm]
-\lambda_{2S} + V & {\displaystyle \frac{\kappa_{2S}}{r}}
\end{array}\right)
\begin{pmatrix}G_{2S}(r)\\[2mm] F_{2S}(r)\end{pmatrix},
\ee
for which $\kappa_{2S} = -1$, $\mu$ again denotes the reduced muon
mass, and $\lambda_{2S}$ denotes the $2S$ eigenvalue shifted by the
reduced mass. In doing so we obtain the upper and lower components
($G_{2S}(r)$ and $F_{2S}(r)$, respectively) of the muon wave-function
in response to the combination, $V$ of the finite-size Coulomb
potential, finite-size vacuum polarization potential, and potential
given by Eq.~(\ref{eq:Vzemach}) (separately for each value of
$\vec{\s}_1\cdot\vec{\s}_2$), for a variety of values of $r_p^C$ and
$r_p^M$.

We express the computed hyperfine splittings (HFS) as a polynomial
function of $r_p^M$ for two significant choices of $r_p^C$\emdash the
2006 CODATA~\cite{Mohr:2008fa} value \mbox{$r_p^C = 0.8768~{\rm fm}$}
(recently updated, as per NIST) and that found in the analysis of
Pohl~\etal~\cite{Pohl:2010zz}, \mbox{$r_p^C = 0.84184~{\rm fm}$}. This
yields
\be \label{eq:abc}
\Delta E_{2S}^{\rm HFS} = A + B(\bra r_p^2\ket_M) + C(\bra r_p^2\ket_M)^{3/2},
\ee
for which the values of $A$, $B$, and $C$ are given in
Table~\ref{tab:abc} for parameterizations at the two values of
$r_p^C$.
\begin{table}[b]
\centering
\caption{\protect\label{tab:abc} Coefficients of Eq.~(\ref{eq:abc})
  parameterizing the $2S$ HFS with magnetic radius, relevant to the
  Zemach correction for two values of the rms charge radius,
  calculated using the finite-Coulomb ($V_{\rm FC}$), finite vacuum
  polarization ($V_{\rm FVP}$), and magnetization potentials ($V_{\rm
    Zemach}$); and alternatively the finite-Coulomb potential plus
  magnetization potential.\vspace{2mm}}
\begin{ruledtabular}
\begin{tabular}{r...}
$r_p^C$  & 
\multicolumn{1}{c}{$A\ [{\rm meV}]$} &
\multicolumn{1}{c}{$B\ [{\rm meV fm}^{-2}]$} & 
\multicolumn{1}{c}{$C\ [{\rm meV fm}^{-3}]$} \\[1mm]
&\\[-4mm]
\hline
&\\[-2mm]
0.84184\footnote{Calculated using $V = V_{\rm FC} + V_{\rm FVP} + V_{\rm Zemach}$}  & 22.6085 & 0.1425 & -0.1191 \\[2mm]
0.8768$^a$   & 22.6037 & 0.1450 & -0.1203 \\[2mm]
0.8768\footnote{Calculated using $V = V_{\rm FC} + V_{\rm Zemach}$}  & 22.5318 & 0.1388 & -0.1142 \\
\end{tabular}
\end{ruledtabular}
\end{table}

\begin{figure*}[!t]
  \includegraphics[angle=90,width=0.7\textwidth]{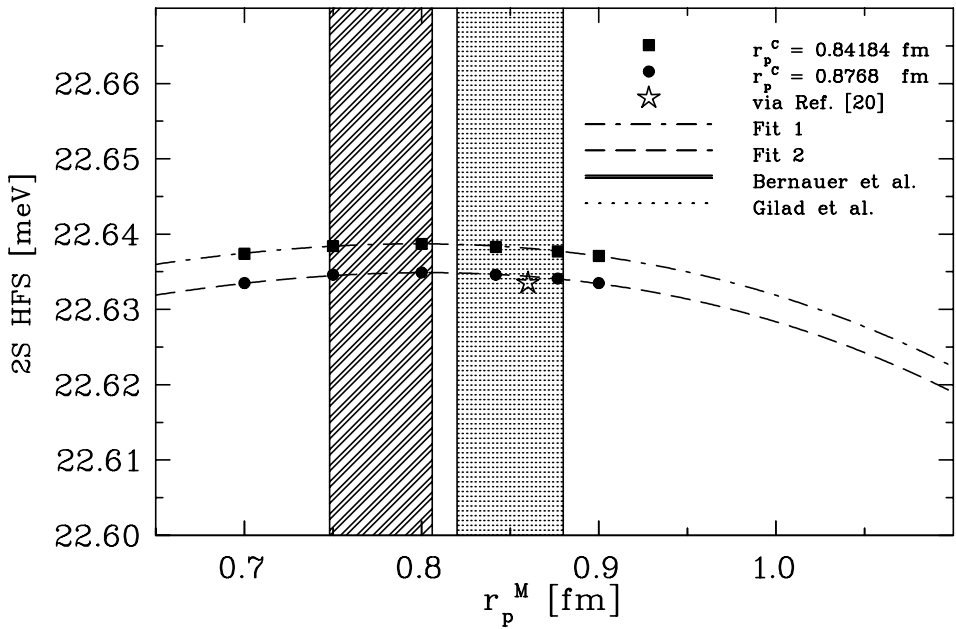}
  \caption{Comparison of data and parameterizations of Zemach
    corrections to the $2S$ hyperfine state of muonic hydrogen for two
    values of the proton rms charge-radius. Also shown are the
    calculated fits (`Fit 1' and `Fit 2' denote fits for $r_p^C =
    0.84184$~fm and $r_p^C = 0.8768$~fm respectively) to the data as
    presented in the first two lines of Table~\ref{tab:abc} and the
    magnetic radii extracted from
    Bernauer~\etal~\cite{Bernauer:2010wm} [Mainz, \mbox{$\bra
        r_M^2\ket^{1/2} = 0.777\pm0.029$~fm}] and
    Gilad~\etal~\cite{Gilad:2011zz} [Jefferson Lab Hall A
      Collaboration, \mbox{$\bra r_M^2\ket^{1/2} = 0.850\pm0.030$~fm}]
    used to calculate the Zemach corrections of
    Eq.~(\ref{eq:corr}). The point denoted by a star corresponds to
    the Zemach correction calculated via charge and magnetization
    densities extracted from Ref.~\cite{Venkat:2010by} at $r_p^C =
    0.878$~fm and $r_p^M = 0.860$~fm as per
    Eq.~(\ref{eq:fitEZ}).\protect\label{fig:zemach}}
\end{figure*}

The two parameterizations are compared in Fig.~\ref{fig:zemach} in
which we note that the dependence on $r_p^M$ is almost non-existent,
while the dependence on $r_p^C$ is stronger. Also shown in
Fig.~\ref{fig:zemach} are the magnetic radii from
Refs.~\cite{Bernauer:2010wm,Gilad:2011zz} corresponding to the
extractions from the Mainz (0.777 $\pm 0.029$ fm) and Jefferson
Laboratory (0.850 $\pm$ 0.030 fm) collaborations, respectively.

The large difference between extracted magnetic radii does not heavily
influence the HFS. The difference between using the Mainz value and
the JLab value is only $0.35$~$\mu$eV for the smaller value of the
charge radius and $0.23$~$\mu$eV for the larger value. Therefore, we
can average the HFS obtained from the two values of the magnetic
radius to calculate the value of the $2S$ hyperfine shift for each of
the value of charge radius. The results are
\be
\label{eq:corr}
\Delta E_{2S}^{\rm HFS, Z} = \left\{
\begin{array}{rl}
22.6384~{\rm meV} &{\ \rm if\ } r_p^C = 0.84184~{\rm fm} \\[2mm]
22.6347~{\rm meV} &{\ \rm if\ } r_p^C = 0.8768~{\rm fm}
\end{array}\right.\, .
\ee

This is the value of the $2S$ hyperfine shift obtained using a
complete treatment of the spatial extent of the proton's
electromagnetic distributions. We can determine the importance of our
complete treatment as compared to using only the Zemach radius by
computing the $2S$ hyperfine shift in the presence of only the
finite-size Coulomb and finite-size vacuum polarization potentials,
which we also fit to a polynomial form to give
\be 
\Delta E_{2S}^{\rm HFS} = 22.8521 
- 0.1795~\bra r_p^2\ket_C 
+ 0.0739~\bra r_p^2\ket_C^{3/2}~{\rm meV},
\ee
as per Ref.~\cite{Carroll:2011rv}. For the two chosen values of
$r_p^C$ this leads to an uncorrected (with respect to the finite size
of the proton magnetization distribution) HFS of
\be
\label{eq:uncorr}
\Delta E_{2S}^{\rm HFS} = \left\{
\begin{array}{rl}
22.7690~{\rm meV} &{\ \rm if\ } r_p^C = 0.84184~{\rm fm} \\[2mm]
22.7639~{\rm meV} &{\ \rm if\ } r_p^C = 0.8768~{\rm fm}
\end{array}\right.\, .
\ee
The isolated Zemach correction calculated here is given by the
difference between the corrected calculation and the uncorrected one,
\bea
\nonumber
\Delta E_{\rm Zemach} &=& \Delta E_{2S}^{\rm
  HFS, Z} - \Delta E_{2S}^{\rm HFS} \\[3mm]
\nonumber
&=& \left\{
\begin{array}{rl}
-0.1306~{\rm meV} &{\ \rm if\ } r_p^C = 0.84184~{\rm fm} \\[2mm]
-0.1292~{\rm meV} &{\ \rm if\ } r_p^C = 0.8768~{\rm fm}
\end{array}\right.\, .\\
&&\label{us}
\eea

In order to identify the significance of vacuum polarization on this
result, we furthermore recalculate the Zemach correction using only
the finite-size Coulomb potential plus potential given by
Eq.~(\ref{eq:Vzemach}). In this case, calculating the potentials using
$r_p^C = 0.8768$~fm for various values of $r_p^M$ we obtain a
polynomial of the form given by Eq.~(\ref{eq:abc}) with values of $A$,
$B$, and $C$ listed in the last line of Table~\ref{tab:abc}.
%
Using this polynomial form we calculate the HFS contribution for the
two values of $r_p^M$ and average (though they are separated by only
twice our numerical error) to obtain
\be
\Delta E_{2S}^{\rm HFS,Z} = 22.5621~{\rm meV}.
\ee

We may compare this to the uncorrected shift calculated using only the
finite-Coulomb potential in Ref.~\cite{Carroll:2011rv}, interpolated
to $r_p^C = 0.8768$~fm, and we obtain
\be
\Delta E_{2S}^{\rm HFS} = 22.6910~{\rm meV},
\ee
indicating a correction using the finite-Coulomb potential alone
(neglecting the vacuum polarization) of 
\be \label{eq:EZFC}
\Delta E_{\rm Zemach} = -0.1289~{\rm meV}.
\ee

The deviation of this value from that in Eq.~(\ref{us}) for the same
value of $r_p^C$ is a mere $0.3~\mu$eV, indicating that vacuum
polarization has negligible influence on the Zemach correction within
our numerical error limits.

Returning to the issue of the significance of the shape of the various
distributions used in these calculations, we have recalculated the
Zemach correction using charge and magnetization densities extracted
from fits to form-factor data~\cite{Venkat:2010by} explicitly. The
so-called charge and magnetization densities $\rho_{C,M}(r)$ may be
extracted from the Sachs electric and magnetic form factors
$G_{E,M}(Q^2)$ via a Fourier transform
\be
\rho_{C,M}(r) \equiv \int \frac{d^3q}{(2\pi)^3}
e^{-i\vec{q}\cdot\vec{r}}G_{E,M}(\vec{q\, }^2)\ ,
\ee
where for an atomic system we have $Q^2=\vec{q\, }^2$, and we take the
Sachs form factors as the fits given in Ref.~\cite{Venkat:2010by} of
the form
\be \label{eq:GEfit}
G_E(Q^2) = 
\frac{1 + q_6\tau + q_{10}\tau^2 + q_{14}\tau^3}
{1 + q_2\tau + q_4\tau^2 + q_8\tau^3 + q_{12}\tau^4 + q_{16}\tau^5}\ ,
\ee
\be \label{eq:GMfit}
G_M(Q^2) = 
\frac{1 + p_6\tau + p_{10}\tau^2 + p_{14}\tau^3}
{1 + p_2\tau + p_4\tau^2 + p_8\tau^3 + p_{12}\tau^4 + p_{16}\tau^5}\ ,
\ee
(noting that the proton magnetic moment $\mu_p$ appears in
Eq.~(\ref{eq:Vzemach}) here rather than in $G_M$) with coefficients
$q_i$, and $p_i$ given in Table~\ref{tab:qp}, and for which $\tau =
Q^2/4M_p^2$. Using the extracted charge density in the finite-Coulomb
potential (which as per Ref.~\cite{Venkat:2010by} is constrained such
that $r_p^C = 0.878$~fm), and the extracted magnetization density
(constrained such that $r_p^M = 0.860$~fm) in the Zemach potential
(the finite vacuum polarization potential, which as noted before is of
minimal consequence, remains calculated using an exponential charge
distribution characteristic of $r_p^C = 0.8768$~fm), we once again
recalculate the Zemach correction to the $2S$ HFS, and we find
\be \label{eq:fitEZ}
\Delta E_{2S}^{\rm HFS,Z} = 22.6335~{\rm meV}\, .
\ee

As we do not have knowledge of either the $r_p^C$ or $r_p^M$
polynomial dependence of this energy, we may only compare to an
interpolation of our previous results at the appropriate
values. Comparing Eq.~(\ref{eq:fitEZ}) to our polynomial fit
corresponding to $r_p^C = 0.8768$~fm (the difference in $r_p^C$ being
minimal) as per the second line of Table~\ref{tab:abc}, and
interpolating to $r_p^M = 0.860$~fm, we find a difference of a mere
$1.0~\mu$eV, indicating that the exponential distribution is indeed a
good approximation for the magnetization density. This Zemach
corrected HFS is included in Figure~\ref{fig:zemach} where it is clear
that it produces good agreement with our usage of a simplified model
of the charge and magnetization densities.


%
\begin{table}[b]
\centering
\caption{\protect\label{tab:qp}Coefficients of polynomial fit to Sachs
  electric and magnetic form factor data for the proton taken
  from~\cite{Venkat:2010by} as used in
  Eqs.~(\ref{eq:GEfit}, \ref{eq:GMfit}).\vspace{2mm}}
\begin{ruledtabular}
\begin{tabular}{r..}
\multicolumn{1}{c}{$i$} & \multicolumn{1}{c}{$p_i$} & \multicolumn{1}{c}{$q_i$} \\[1mm]
&\\[-4mm]
\hline
& \\[-2mm]
 2 & 9.70703681 & 14.5187212 \\[2mm]
 4 & 0.00037357 & 40.88333 \\[2mm]
 6 & -1.43573   & 2.90966 \\[2mm]
 8 & 0.00000006 & 99.999998 \\[2mm]
10 & 1.19052066 & -1.11542229 \\[2mm]
12 & 9.9527277  & 0.00004579 \\[2mm]
14 & 0.25455841 & 0.003866171 \\[2mm]
16 & 12.7977739 & 10.3580447 \\
\end{tabular}
\end{ruledtabular}
\end{table}

We compare our values of the Zemach correction (Eq.~(\ref{us}) and
Eq.~(\ref{eq:EZFC})) to that which was used in the analysis of
Pohl~\etal~\cite{Pohl:2010zz}, {\it viz.} that of
Martynenko~\cite{Martynenko:2004bt} (denoted in that reference as
`Proton structure corrections of order $\alpha^5$')
\bea
\Delta E_{\rm Zemach}^{\rm Martynenko} = -0.1518~{\rm meV}.
\label{marty}\eea
This includes the influence of the Zemach term as well as the small
($\sim 5$ \%, depending on the value of the Zemach radius) effects of
a putative treatment of the two-photon exchange interaction (with
intermediate nucleon states).

\newcommand{\eq}[1]{Eq.~(\ref{#1})} The comparison between \eq{us} and
\eq{marty} suggests that the value used in the analysis of
Pohl~\etal~requires a significant modification of some $10$--$15\%$.
A more precise consideration of the importance of using a complete
non-perturbative treatment is to look at the the next-leading
contribution of Ref.~\cite{Martynenko:2004bt} (`Proton structure
correction of order $\alpha^6$' = -0.0017~meV). This is an order of
magnitude smaller than the difference found here, indicating that the
perturbative approach is not satisfactory in this scenario.

We may also study the importance of the Zemach radius by examining the
values of the Zemach radii for the values of $r_p^{C,M}$ of present
interest, as displayed in Table~\ref{tab:zem}. The HFS is almost
independent of the value of $r_p^M$, but the Zemach radius displays a
very significant dependence on this parameter.

\begin{table}[t]
\centering
\caption{\protect\label{tab:zem} Comparison of Zemach radii
  (\eq{eq:EZzemach}) for several choices of $r_p^{C}$ and values of
  $r_p^M$ taken from Refs.~\cite{Bernauer:2010wm} \mbox{($r_p^M =
    0.777$~fm)}, \cite{Gilad:2011zz} \mbox{($r_p^M = 0.850$~fm)}, and
  \cite{Venkat:2010by} \mbox{($r_p^M = 0.860$~fm)}.\vspace{2mm}}
\begin{ruledtabular}
\begin{tabular}{r....}
$r_p^C$ & 
\multicolumn{1}{c}{$R_p$ (fm) \cite{Bernauer:2010wm}} &
\multicolumn{1}{c}{$R_p$ (fm) \cite{Gilad:2011zz}} & 
\multicolumn{1}{c}{$R_p$ (fm) \cite{Venkat:2010by}} 
\\[1mm]
&\\[-4mm]
\hline
&\\[-2mm]
0.84184  & 1.022 & 1.068 & \\[2mm]
0.8768   &  1.046 & 1.091 & \\[2mm]
0.878    & & & 1.081
\end{tabular}
\end{ruledtabular}
\end{table}

The difference between our value quoted in~\eq{us} and the value of
Martynenko quoted in~\eq{marty} does not resolve the proton radius
puzzle. The effect presented here alters the $2S$
hyperfine splitting by $0.02$~meV, contributing only $0.005$~meV
(while 0.3 meV is needed) to the transition measured by
Ref.~\cite{Pohl:2010zz}. However, our results add weight to the
argument that the perturbative corrections require further attention
and re-evaluation using the latest numerical tools, along with a close
examination of the relevant physics to ensure that no contribution is
overlooked, such as that reported in Ref.~\cite{Miller:2011yw}.

Recomputing the $2S$ HFS with this new Zemach inclusion, we obtain a
splitting as given in Table~\ref{tab:2SHFS}, which summarizes our
updated non-perturbative calculation of this splitting.

We note a relevant error in Ref.~\cite{Carroll:2011rv}\emdash the
Zemach contributions to the $2S$ HFS labelled `Proton structure
corrections of order $\alpha^5$ and $\alpha^6$' are listed in Table
III of Ref.~\cite{Carroll:2011rv} as included by the Dirac
calculation, whereas these terms should appear under `Remaining
corrections'. This contribution now appears corrected in
Table~\ref{tab:2SHFS} which should serve as a replacement to the
former table. The re-calculation of the Zemach contribution alters the
prediction of the proton rms radius via the analysis of
Ref.~\cite{Carroll:2011rv} to arrive at
\be
r_p^C = 0.84182(67)~{\rm fm},
\ee
which does not differ from the former conclusion (\mbox{$r_p^C =
  0.83811(67)~{\rm fm}$}) in a statistically significant fashion with
regard to the main discrepancy, but does agree with that of 
Pohl~\etal~to a high level of precision.

We conclude by stating that our calculation of the influence of the
spatial extent of the proton's electromagnetic distributions using a
non-perturbative, relativistic framework results in a 10--15$\%$
deviation of the Zemach correction from the value used by
Pohl~\etal. Furthermore, the very weak dependence of the HFS on the
value of the magnetic radius, shows that the perturbative treatment,
dominated by the influence of the Zemach radius, does not capture the
effects of the proton's spatial distributions on the HFS. This result
invites and encourages further investigation into the many corrections
which enter the analysis of the muonic hydrogen spectrum at the level
of precision required to extract the value of the proton radius at the
required level of precision.

\begin{center}
\begin{table*}[t]
\caption{Contributions to the 2S$_{1/2}$ hyperfine splitting with
  comparison to values found in Martynenko~\cite{Martynenko:2004bt}
  corrected and updated from Ref.~\cite{Carroll:2011rv} to include the
  re-calculated Zemach contribution. Values are all in meV. Errors in
  the Dirac calculations are taken to be $\pm 500$~neV. The listed
  corrections are already included in our Dirac calculations and are
  listed by their descriptions in Ref.~\cite{Martynenko:2004bt}. All
  further corrections to both the perturbative calculation and our
  calculation are contained in `Remaining Corrections' which
  encompasses the muon AMM, amongst other corrections listed in
  Ref.~\cite{Martynenko:2004bt}. We note that the in
  Ref.~\cite{Martynenko:2004bt} the Zemach correction is listed as
  `Proton structure corrections of ${\cal O}(\alpha^5)$' and may not
  include considerations of finite-size in the wave-function. The
  polynomial dependence on $\bra r_p^2\ket^n$ of this splitting is not
  discussed in the literature. The strict $r_p^C$ dependence of the
  Zemach correction has not been calculated, and as such we have
  selected a value (that corresponding to $r_p^C = 0.8768~{\rm fm}$)
  suitable to our previous
  analysis~\cite{Carroll:2011rv}. \label{tab:2SHFS}\vspace{2mm}}
\begin{tabular}{l.c}
&& \\[-3mm]
\hline
\hline\\[-2mm]
Contribution & \multicolumn{1}{c}{\phantom{abc}Martynenko\phantom{abc}} & Present Work \\[1mm]
\hline\\[-2mm]

Dirac ($V = V_{\rm C}$) & & 22.8229 \\[1mm]
Dirac ($V = V_{\rm C} + V_{\rm VP}$) & & 22.8976 \\[1mm]
Dirac ($V = V_{\rm FC}$) & & 22.7774 - 0.1746 $\bra r_p^2\ket$ + 0.0709 $\bra r_p^2\ket^{3/2}$ \\[1mm]
Dirac ($V = V_{\rm FC} + V_{\rm VP}$) & & 22.8510 - 0.1701 $\bra r_p^2\ket$ + 0.0667 $\bra r_p^2\ket^{3/2}$ \\[1mm]
\hline
&& \\[-2mm]
Dirac ($V = V_{\rm FC} + V_{\rm FVP}$) & & 22.8521 - 0.1795 $\bra r_p^2\ket$ + 0.0739 $\bra r_p^2\ket^{3/2}$ \\[1mm]

Fermi Energy $E_F$ &
22.8054 &   \\[1mm]

Relativistic correction $\frac{17}{8}(Z\alpha)^2E_F$ & 
0.0026 &   \\[1mm]

VP corrections of orders $\alpha^5$, $\alpha^6$ & & \\[1mm]
\ \ in the second order of perturbation series &
0.0746 & \\[1mm]

Proton structure corrections of order $\alpha^5$ &
-0.1518 &   \\[1mm]

Proton structure corrections of order $\alpha^6$ &
-0.0017 &   \\[1mm]

Zemach correction &
 & -0.1292  \\[1mm]

\hline
&& \\[-2mm]
Subtotal: & 22.7291 & 22.7229 - 0.1795 $\bra
r_p^2\ket$ + 0.0739 $\bra r_p^2\ket^{3/2}$ \\[1mm]

\hline
&& \\[-2mm]

Remaining Corrections & 0.0857 & 0.0857 \\[1mm]

\hline
&& \\[-2mm]
Total: & \multicolumn{1}{c}{22.8148~$\pm$~0.0078\qquad} & 22.8086 - 0.1795 $\bra
r_p^2\ket$ + 0.0739 $\bra r_p^2\ket^{3/2}$ \\[1mm]

\hline
\hline

\end{tabular}
\end{table*}
\end{center}

\begin{acknowledgments}
\vspace{-4mm}

This research was supported in part by the United States Department of
Energy (under which Jefferson Science Associates, LLC, operates
Jefferson Lab) via contract DE-AC05-06OR23177 (JDC, in part); grants
FG02-97ER41014 and LBNL DE-AC02-05CH11231(GAM); and grant
DE-FG02-04ER41318 (JR), and by the Australian Research Council
(through grant FL0992247) and the University of Adelaide (JDC,
AWT). GAM and JR gratefully acknowledge the support and hospitality of
the University of Adelaide while the project was undertaken.

\end{acknowledgments}
%
%
\bibliography{muHrefs}
\end{document}